\documentclass[prl,amsmath,amssymb,superscriptaddress,twocolumn]{revtex4}
\usepackage{graphicx}
\usepackage{bm}

\newcommand{\bk}{{\bf k}}

\newcommand{\bn}{{\bf n}}

\newcommand{\bp}{{\bf p}}

\newcommand{\bQ}{{\bf Q}}

\newcommand{\eps}{\epsilon}

\DeclareMathAlphabet{\mathpzc}{OT1}{pzc}{m}{it} 
\begin{document}
\title{Spin-orbit coupling induced enhancement of superconductivity in a two-dimensional repulsive gas of fermions}
\author{Oskar Vafek}
\author{Luyang Wang}
%\altaffiliation[Also at ]{Department of Physics, FSU}
\affiliation{National High Magnetic Field Laboratory and Department
of Physics, Florida State University, Tallahassee, Florida 32306,
USA}
\address{}

\date{\today}
\begin{abstract}
We study a model of a two-dimensional repulsive Fermi gas with
Rashba spin-orbit coupling $\alpha_R$, and investigate the
superconducting instability using renormalization group approach. We
find that in general superconductivity is enhanced as the
dimensionless ratio $\frac{1}{2}m\alpha_R^2/E_F$ increases,
resulting in unconventional superconducting states which break time
reversal symmetry.
\end{abstract}
\maketitle

There is a growing interest in materials whose interfaces support a
two-dimensional (2D) electron gas and display superconductivity,
because of their novel, and potentially technologically useful
properties such as electronic transport, magnetism and interplay
between structural
instabilities\cite{TrisconeReview,HarroldHwang,LuLi}. Due to the
intrinsic breaking of the inversion symmetry, spin-orbit coupling is
expected to play a role in determining the nature of the
superconducting state. For example, experimentally the enhancement
of transition temperature at LaAlO$_3$/SrTiO$_3$ interfaces tracks
the enhancement of Rashba spin-orbit coupling\cite{TrisconePRL}. And
while the mechanism of superconductivity here is likely related to
the electron-phonon mechanism of the bulk materials\cite{CohenPR},
such considerations motivate us to investigate the effect of
spin-orbit coupling on the superconducting transition.

For attractive interactions the question has been addressed in
Ref.\cite{Edelshtein,GorkovRashba,MineevSigrist}. In contrast, here
we consider a model of {\it repulsive} fermions moving in 2D and
analyze the nature of the unconventional superconducting state in
weak coupling. For a strictly parabolic dispersion in 2D, without
spin-orbit coupling, it is known that repulsive interactions do not
induce superconductivity to second order in the interaction, unlike
in 3D where $p$-wave superconductivity is found at this order. In 2D
one has to go to third order\cite{ChubukovPRB} for the
Kohn-Luttinger effects to appear. Our motivation is to understand
the role of the spin-orbit coupling in this process, to determine
whether it can enhance superconductivity, and to study the nature of
the superconducting state. Since we treat the Rashba spin-orbit
coupling $\alpha_R$ non-perturbatively, we can analyze the relative
values of the mean-field transition temperatures $T_{c}$ for an
arbitrary value of the dimensionless ratio
$\Theta=\frac{1}{2}m\alpha^2_R/E_F$, where $m$ is the (bare) fermion
mass and $E_F$ is the Fermi energy, measured from the Dirac point
(see Fig. 1). In the strictest sense the transition in 2D is of
Kosterlitz-Thouless type and at $T_{KT}<T_{c}$. However, since we
are working in the weak coupling limit, the pairing energy scale is
much smaller than the zero temperature phase stiffness energy and
$1-T_{KT}/T_{c}\sim T_{c}/E_F(1+\Theta)\ll 1$, justifying the
approach presented here.

Due to the spin-orbit interaction, the pair states cannot be chosen
to be pure spin singlet or triplet, but appear as linear
superposition thereof\cite{GorkovRashba}. Nevertheless, since the
Rashba model (\ref{eq:Hkin}), as well as the short range repulsion
(\ref{eq:Hint}), commute with the $z-$component of the total angular
momentum $J_z=L_z+S_z$, we can label the pair states according to
$\ell$, the eigenvalue of $J_z$. For small values of $\Theta$ we
find that states with high values of relative angular momentum
$\ell$ condense first, with $\ell$ decreasing as $\Theta$ increases.
For intermediate values of $\Theta$ we find broad regions of
stability for $\ell=4$, with dome-like dependence of $T_c$ on
$\Theta$, while in the limit of large $\Theta$, we find $\ell=2$. In
weak coupling we show that all of these states spontaneously break
time-reversal symmetry. While we formulate our calculation within
more modern renormalization group (RG) approach, our results can be
rederived diagrammatically by summing the leading logarithms to all
orders in perturbation theory, as has been done traditionally in
treating Kohn-Luttinger
effect\cite{GorkovMelikBarkhudarov,ChubukovReview}. Also, while our
approach is similar to that of Ref.\cite{RaghuKivelsonScalapino}
(see also\cite{RaghuKivelson2011}), we use a single step RG instead
of a two step RG, which we find more economical.

\begin{figure}[t]
\begin{center}
\begin{tabular}{cc}
\includegraphics[width=0.2\textwidth]{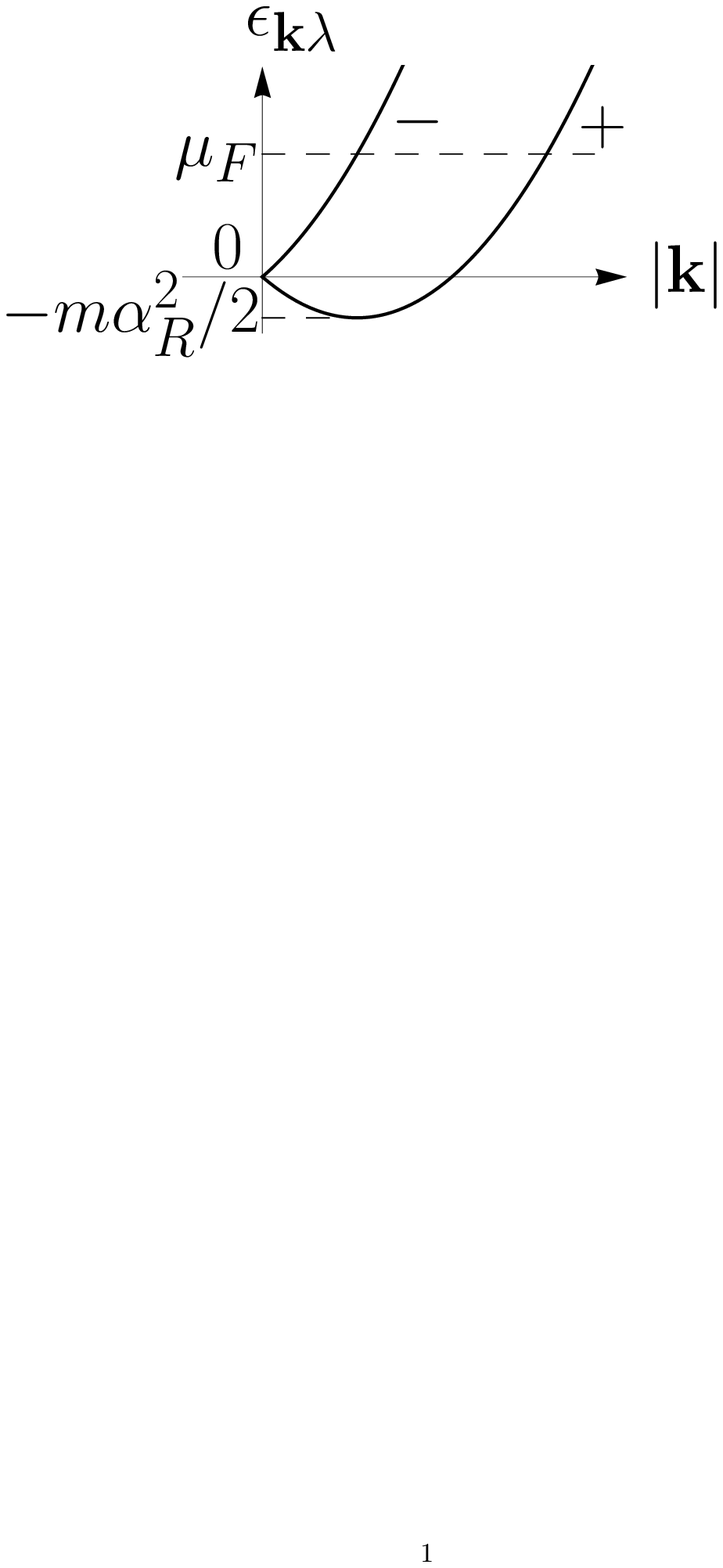}&
\includegraphics[width=0.1\textwidth]{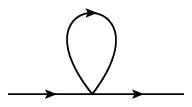}
\end{tabular}
\end{center}
 \caption{(Left) The dispersion relation.
 (Right) First order (tadpole) correction to self-energy.
 }\label{fig:dispersion}
\end{figure}

 Our starting point is the Hamiltonian for
Fermions moving in 2D
\begin{eqnarray}\label{eq:Htot}
\mathcal{H}&=&H_{kin}+H_{int}
\end{eqnarray}
where in momentum representation the kinetic energy (including
spin-orbit coupling) is
\begin{eqnarray}\label{eq:Hkin}
H_{kin}&=&\sum_{\bk,\alpha\beta}c^{\dagger}_{\bk\alpha}\left(\frac{\bk^2}{2m}\delta_{\alpha\beta}+\alpha_R({\bm\sigma}_{\alpha\beta}\times\bk)\cdot
\hat{\bn} \right)c_{\bk\beta}
\end{eqnarray}
and the short-range interaction energy term is
\begin{eqnarray}\label{eq:Hint}
H_{int}&=&\frac{u}{2}\frac{1}{L^2}\!\!\!\!\!\sum_{\bk_1\ldots
\bk_4,\sigma\sigma'}\!\!\!\!\!\delta_{\bk_1+\bk_2,\bk_3+\bk_4}
c^{\dagger}_{\bk_1\sigma}c^{\dagger}_{\bk_2\sigma'}c_{\bk_3\sigma'}c_{\bk_4\sigma}.
\end{eqnarray}
As usual, the components of $\bk$ belong to
the Born-von Karman set $\{2\pi n/L\}$ where $n$ is an integer and
$L$ is the linear size of the system. Unlike in
Ref.\cite{GorkovRashba}, we consider superconductivity for repulsive
interactions, i.e. for $u>0$, in the weak coupling limit
$u\nu_{2D}<<1$, where the density of states per spin in 2D for
$\alpha_R=0$ is $\nu_{2D}=\frac{m}{2\pi}$. The kinetic energy term
is diagonalized using the following transformation
\begin{eqnarray} \left(\begin{array}{c}
c_{\bk\uparrow}\\
c_{\bk\downarrow}
\end{array}\right)
= \frac{1}{\sqrt{2}}\left(
\begin{array}{cc}
1 & 1\\
ie^{i\phi_{\bk}} & -ie^{i\phi_{\bk}}
\end{array}\right)
\left(\begin{array}{c}
a_{\bk +}\\
a_{\bk -}
\end{array}\right).
\end{eqnarray}
Next, we rewrite the Hamiltonian in terms of these helicity
eigenmodes. The partition function associated with $\mathcal{H}$ can
be expressed in terms of the coherent state Feynman path integral
over Grassman variables\cite{NegeleOrland} as
\begin{eqnarray}
Z=\int\mathcal{D}[a^*_{\lambda}(\tau)a_{\lambda}(\tau)]e^{-S_0-S_{int}}
\end{eqnarray}
where
\begin{eqnarray}
S_0&=&\int_0^{\beta}d\tau
\sum_{\bk,\lambda=\pm}a^*_{\bk\lambda}(\tau)\left(\frac{\partial}{\partial
\tau}+\eps_{\bk\lambda}-\mu_F\right)a_{\bk\lambda}(\tau),\nonumber\\
S_{int}&=&\sum_{1,2,3,4}U(1,2,3,4)a^*(1)a^*(2)a(3)a(4),
\end{eqnarray}
where the single particle energies are (see Fig.\ref{fig:dispersion})
\begin{eqnarray}
\eps_{\bk\lambda}=\frac{\bk^2}{2m}-\lambda\alpha_R k.
\end{eqnarray}
In the above expressions $\beta=1/(k_BT)$, $\mu_F$ is the {\it
exact} chemical potential whose value depends on temperature $T$ and
interaction $u$, in such a way as to preserve average particle
density. We adopt a shorthand expression for the multiple summations
$\sum_{1,2,3,4}(\ldots)\equiv\int_0^{\beta}d\tau_1\ldots d\tau_4
\sum_{\bk_1\ldots\bk_4}\sum_{\mu\nu\lambda\rho}(\ldots)$,
\begin{eqnarray}
&&U(1,2,3,4)=-\frac{u}{16L^2}\int_0^{\beta}d\tau\prod_{j=1}^4\delta(\tau-\tau_j)\delta_{\bk_1+\bk_2,\bk_3+\bk_4}\nonumber\\
&&\times\left(\mu e^{-i\phi_{\bk_1}}-\nu
e^{-i\phi_{\bk_2}}\right)\left(\lambda e^{i\phi_{\bk_3}}-\rho
e^{i\phi_{\bk_4}}\right),
\end{eqnarray} and
$a(j)=a_{\bk_j\alpha_j}(\tau_{j})$ where
$\alpha_j=\{\mu,\nu,\lambda,\rho\}$ and $\phi_{\bk}$ is an azimuthal
angle in the momentum plane.

We proceed by integrating out the high energy modes between the
energy cutoff $A$ and $\Omega\ll A$ about the two Fermi surfaces at
$T=0$. The expansion is organized by the powers of the dimensionless
parameters $u\nu_{2D}$ and $\Omega/A$. At first order in the
cumulant expansion, we find a correction to the chemical potential
$\mu_F$ from the tadpole diagram shown in Fig.\ref{fig:dispersion}.
This correction is $\delta
\mu_{F}=-\frac{1}{2}u(\langle\hat{\rho}_{+}\rangle+\langle\hat{\rho}_{-}\rangle)$,
where
$\hat{\rho}_{\pm}=\int\frac{d^2\bk}{(2\pi)^2}a^{\dagger}_{\bk\pm}a_{\bk\pm}$.
Such {\it negative} interaction correction must be absorbed in the
chemical potential counterterm,
$\mu_F-\mu^{(0)}_F=\frac{1}{2}u(\langle\hat{\rho}_{+}\rangle+\langle\hat{\rho}_{-}\rangle)+\mathcal{O}(u^2)$,
which is {\it positive}, and which guarantees that the average
particle density remains fixed. In general, we are not aware of any
argument why interactions should not renormalize the areas of the
individual Fermi surfaces, while of course maintaining their sum
fixed, but to first order we find no such renormalization.
\begin{figure}[t]
\begin{center}
\begin{tabular}{cc}
\includegraphics[width=0.18\textwidth]{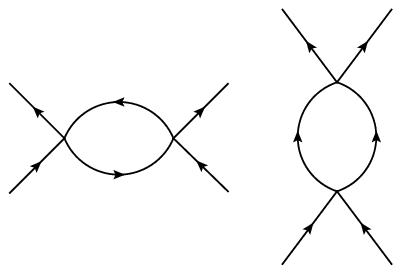}&
\includegraphics[width=0.25\textwidth]{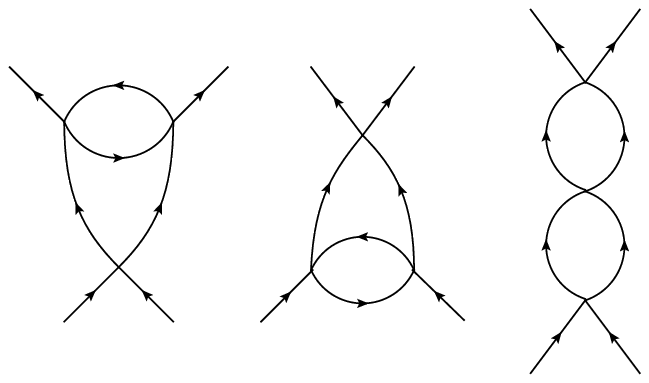}
\end{tabular}
\includegraphics[width=0.35\textwidth]{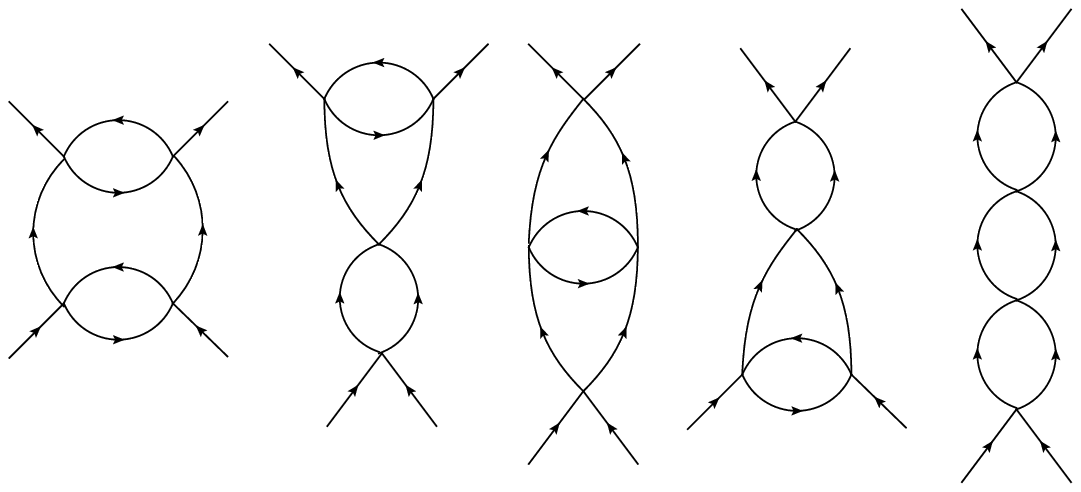}
\end{center}
  \caption{(First row) 2$^{nd}$ and 3$^{rd}$ order corrections to the 4-pt scattering
  amplitude.
  (Second row) 4$^{th}$ order correction. For the 3$^{rd}$ and
  4$^{th}$ order terms, we display only the diagrams which contain
  logarithmic enhancement.
  }\label{fig:graphs}
\end{figure}

Superconducting instability comes from second and higher order terms
in cumulant expansion. We first find the renormalization of the
general four fermion term and then we place the pairs on the two
Fermi surfaces, which are the only processes with logarithmic
enhancements. To second order in $u$, and in the Cooper channel, we
have the following correction to the effective interaction action
$\delta S_{int}=$
\begin{eqnarray}
\frac{u^2}{64L^2}\int_0^{\beta}\!\!d\tau\sum_{\bk\bk'}\sum_{\mu\lambda}
V_{\mu\lambda}(\bk,\bk')a^{*}_{\bk\mu}(\tau)a^{*}_{-\bk\mu}(\tau)a_{-\bk'\lambda}(\tau)a_{\bk'\lambda}(\tau)\nonumber
\end{eqnarray}
where the sum over $\bk,\bk'$ is restricted to a small window near
the Fermi surfaces defined by indices $\mu$ and $\lambda$ within the
energy $\Omega$ above and below $\mu_F$. We write
\begin{eqnarray}
V_{\mu\lambda}(\bk,\bk')=V^{pp}_{\mu\lambda}(\bk,\bk')+V^{ph}_{\mu\lambda}(\bk,\bk'),
\end{eqnarray}
where the two qualitatively different contributions, arising from
the two 2$^{nd}$ order diagrams shown in Fig.\ref{fig:graphs} are
\begin{eqnarray}\label{eq:Vpp}
V^{pp}_{\mu\lambda}(\bk,\bk')&=&-8\mu\lambda(N_++N_-)e^{-i\phi_{\bk}}e^{i\phi_{\bk'}}\ln\frac{A}{\Omega}\\
\label{eq:Vph}
V^{ph}_{\mu\lambda}(\bk,\bk')&=&\Pi_{\mu\lambda}(\bk,\bk')-\Pi_{\mu\lambda}(-\bk,\bk').
\end{eqnarray}
The density of states on the two Fermi surfaces are
$N_{\pm}=\nu_{2D}\left(1\pm\frac{\sqrt{\Theta}}{\sqrt{1+\Theta}}\right)$.
In the second "particle-hole" contribution
\begin{eqnarray}\label{eq:Pi1}
&&\Pi_{\mu\lambda}(\bk,\bk')=\sum_{\alpha,\beta=\pm}\int\frac{d^2\bp}{(2\pi)^2}
\frac{n_F(\eps_{\bp\alpha})-n_F(\eps_{\bp+\bk-\bk'\beta})}{\eps_{\bp\alpha}-\eps_{\bp+\bk-\bk'\beta}}\nonumber\\
&\times& \left(\mu e^{-i\phi_{-\bk}}-\beta
e^{-i\phi_{\bp+\bk-\bk'}}\right) \left(\lambda
e^{i\phi_{-\bk'}}-\alpha e^{i\phi_{\bp}}\right)
\nonumber\\
&\times&\left(\alpha e^{-i\phi_{\bp}}-\mu e^{-i\phi_{\bk}}\right)
\left(\beta e^{i\phi_{\bp+\bk-\bk'}}-\lambda
e^{i\phi_{\bk'}}\right),
\end{eqnarray}
where the Fermi occupation factor
$n_F(x)=1/(e^{(x-\mu^{(0)}_F)/T}+1)$, evaluated in the limit
$T\rightarrow 0$. After a somewhat tedious, but otherwise
straightforward analysis we find that we can write
\begin{eqnarray}\label{eq:Pi2}
\Pi_{\mu\lambda}(\bk,\bk')&=&2m
e^{-i\phi_{\bk}}e^{i\phi_{\bk'}}\Lambda_{\mu\lambda}(\Theta,\cos(\phi_{\bk}-\phi_{\bk'}))
\end{eqnarray}
where $\Lambda_{\mu\lambda}(\Theta,\cos(\phi_{\bk}-\phi_{\bk}'))$ is
real. Note that under time reversal the helicity basis creation and
annihilation operators transform as $\hat{K}a_{\bk\pm}=\mp
ie^{i\phi_{\bk}}a_{-\bk\pm}$ and $\hat{K}a^{\dagger}_{\bk\pm}=\pm
ie^{-i\phi_{\bk}}a^{\dagger}_{-\bk\pm}$ respectively, where we used
$\phi_{-\bk}=\phi_{\bk}+\pi$. The above relation means that the
Cooper channel potential $V_{\mu\lambda}(\bk,\bk')$ pairs time
reversed states, as it should\cite{MineevSigrist}. Inspecting the
form of the remaining terms in (\ref{eq:Vpp}) as well as the
combination
$\Lambda^{(S)}_{\mu\lambda}(\Theta,\cos\phi)=\frac{1}{2}\Lambda_{\mu\lambda}(\Theta,\cos\phi)+\frac{1}{2}\Lambda_{\mu\lambda}(\Theta,-\cos\phi)$
appearing in (\ref{eq:Vph}), shows that they are invariant under
operations of the 2D rotation group. Additionally, since the {\it
remaining} terms in the scattering amplitude are even under
$\bk\rightarrow -\bk$, and independently under $\bk'\rightarrow
-\bk'$, they can be decomposed into sum over even angular momentum
channels
\begin{eqnarray}\label{eq:Vell}
V^{ph}_{\mu\lambda}(\bk,\bk')=4m
e^{-i\phi_{\bk}}e^{i\phi_{\bk'}}\!\!\!\sum_{\ell=0,2,4,\ldots}\!\!\!V^{(\ell)}_{\mu\lambda}\cos(\ell(\phi_{\bk}-\phi_{\bk'}))
\end{eqnarray}
where the dimensionless Fourier coefficients
$V^{(\ell)}_{\mu\lambda}$ are functions of $\Theta$ and represent
intra- and inter-band pairing amplitudes.

\begin{figure}[t]
\begin{center}
%\begin{tabular}{cc}
\includegraphics[width=0.34\textwidth]{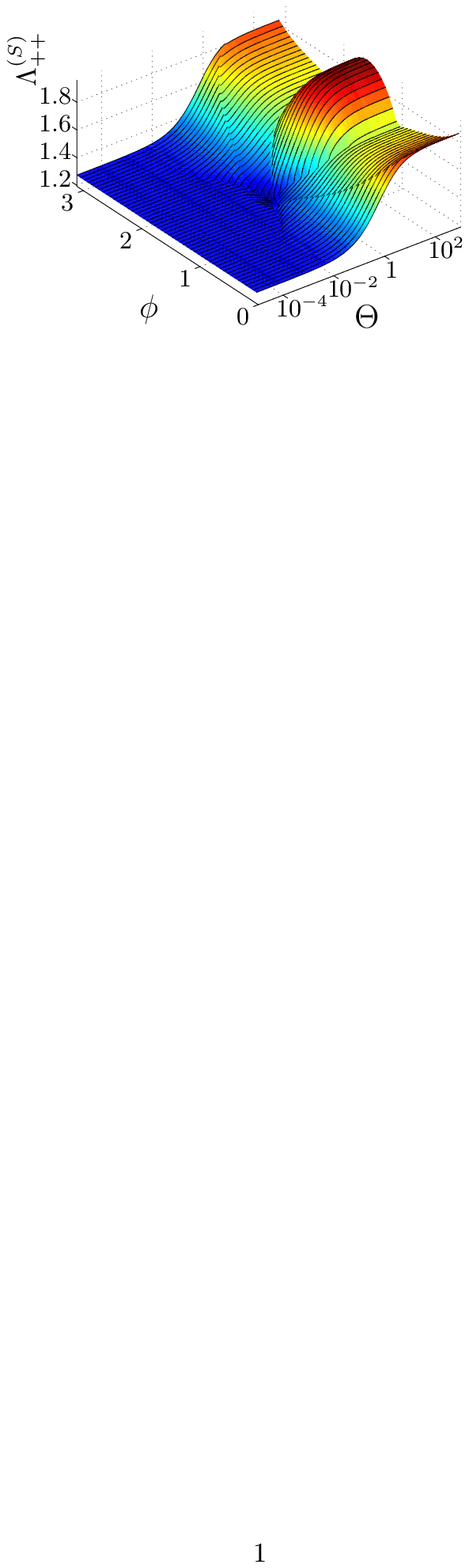}
\includegraphics[width=0.34\textwidth]{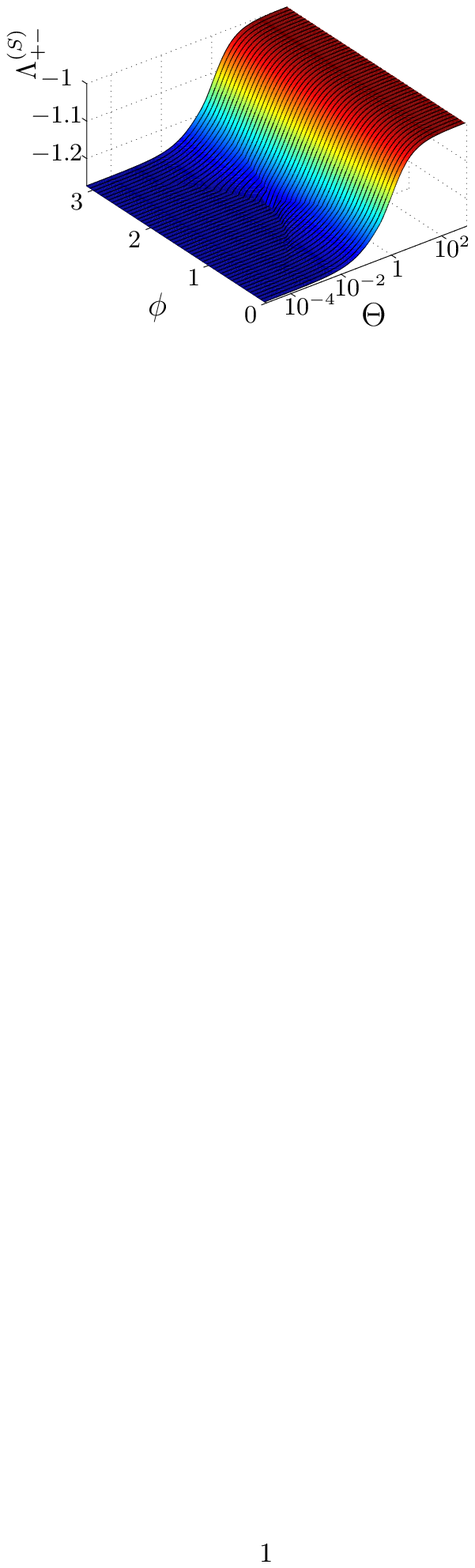}
%\end{tabular}
\end{center}
 \caption{Relative angle $\phi=\phi_{\bk}-\phi_{\bk'}$ and $\Theta=\frac{1}{2}m\alpha^2_R/E_F$ dependence of the interaction function $\Lambda^{(S)}_{\mu\nu}$
 Eqs.(\ref{eq:Vph}-\ref{eq:Pi2}) and text below.
 $\Lambda^{(S)}_{++}$ (top) and $\Lambda^{(S)}_{+-}$ (bottom) start from
 $\pm\frac{4}{\pi}$ at $\Theta=0$ and develop $\phi$ dependence for
 finite $\Theta$, while $\Lambda^{(S)}_{--}$ remains $\frac{4}{\pi}$
 for any $\Theta$.
 }\label{fig:Lambda}
\end{figure}

In order to determine $V^{(\ell)}_{\mu\lambda}$, we need to evaluate
$\Lambda_{\mu\lambda}(\Theta,\cos\phi)$ in Eq.(\ref{eq:Pi2}) from
Eq.(\ref{eq:Pi1}). We shift $\bp\rightarrow \bp-\frac{1}{2}\bQ$
where $\bQ=\bk-\bk'$, and transform from the polar coordinates to
elliptical coordinates $x\in [1,\infty)$, $\psi\in[0,2\pi)$ by
substituting $p_{\parallel}=\frac{1}{2}|\bQ|x\cos\psi$ and
$p_{\perp}=\frac{1}{2}|\bQ|\sqrt{x^2-1}\sin\psi$. In the resulting
expression $\psi$ appears only in $\cos\psi$, so we can substitute
$y=\cos\psi$. For $\alpha=\beta$ we then perform the integral over
$y$ first, which can be done in terms of elementary functions.
Similarly, for $\alpha=-\beta$ we perform the integral over $x$
first. Our analysis is based on numerical integration of the
remaining integral, which can be done quite fast to any desired
accuracy. The final result for the antisymmetrized combination
$\Lambda^{(S)}_{\mu\lambda}(\Theta,\cos\phi)$ is shown in the
Fig.\ref{fig:Lambda}.

Next, we consider 3rd and 4th order terms in $u$ which renormalize
the Cooper channel. These terms can be represented by diagrams shown
in Fig \ref{fig:graphs}, and used to derive the RG equations
governing the flows of Cooper channel couplings, which decouple in
the angular momentum basis. For $\ell\neq 0$ we find that the
renormalized coupling
\begin{eqnarray}\label{eq:Vren ell neq0}
{V^{r}}^{(\ell)}_{\mu\lambda}=\frac{u^2m}{2^5}V^{(\ell)}_{\mu\lambda}
-\frac{u^4m^2}{2^9}\sum_{\alpha=\pm}N_{\alpha}V^{(\ell)}_{\mu\alpha}V^{(\ell)}_{\alpha\lambda}\ln\frac{A}{\Omega}+\ldots
\end{eqnarray}
where $\ldots$ represents term of order $u^4$ which do not contain
(large) logarithm as well as terms of higher order in $u$. If we
define a dimensionless coupling matrix
$g^{(\ell)}_{\mu\lambda}=\frac{1}{2^5}u^2m\sqrt{N_{\mu}N_{\lambda}}V^{(\ell)}_{\mu\lambda}$
and take the logarithmic derivative of the right hand side in
(\ref{eq:Vren ell neq0}), then to, and including,
$\mathcal{O}(u^4)$, we find
\begin{eqnarray}\label{eq:Vren ell neq0 RG}
\frac{d{g^{r}}^{(\ell)}_{\mu\lambda}}{d\ln\Omega}&=&2\sum_{\alpha=\pm}{g^{r}}^{(\ell)}_{\mu\alpha}{g^{r}}^{(\ell)}_{\alpha\lambda}.
\end{eqnarray}
As usual, we have replaced the bare couplings by renormalized
couplings to the order we are working. For $\ell\neq 0$, the initial
condition for the above (matrix) differential equation is
${g^{r}}^{(\ell)}_{\mu\lambda}|_{\Omega=A}=\frac{1}{2^5}u^2m\sqrt{N_{\mu}N_{\lambda}}V^{(\ell)}_{\mu\lambda}$.
This equation can be readily integrated by transforming into the
orthonormalized basis for ${g^{r}}^{(\ell)}_{\mu\lambda}(\Omega)$
with eigenvalues
\begin{eqnarray}\label{eq:gren flow}
{g^{r}}^{(\ell)}_{\pm}(\Omega)=\frac{{g}^{(\ell)}_{\pm}}{1+2{g}^{(\ell)}_{\pm}\ln\frac{A}{\Omega}}
\end{eqnarray}
where the initial eigenvalues of
${g^{r}}^{(\ell)}_{\mu\lambda}|_{\Omega=A}$, for $\ell\neq0$, are
\begin{eqnarray}\label{eq:gell pm}
{g}^{(\ell)}_{\pm}&=&\frac{u^2m}{2^5}\left(\frac{1}{2}(N_+V^{(\ell)}_{++}+N_-V^{(\ell)}_{--})\nonumber\right.\\
&\pm&\left.\sqrt{\frac{1}{4}(N_+V^{(\ell)}_{++}-N_-V^{(\ell)}_{--})^2+N_+N_-{V^{(\ell)}}^2_{+-}}\right)
\end{eqnarray}
If ${g}^{(\ell)}_{\pm}<0$ for some $\ell$ or $\Theta$, then the
associated renormalized coupling (\ref{eq:gren flow}) diverges at a
scale
\begin{eqnarray}
T^{(\ell)}_c\sim {\Omega^*}^{(\ell)}=Ae^{-1/|g^{(\ell)}_{eff,\pm}|}
\end{eqnarray}
where $g^{(\ell)}_{eff,\pm}=2g^{(\ell)}_{\pm}$. While the assignment
between $T_c$ and $\Omega^*$ cannot reliably determine the prefactor
of the exponential term, the {\it relative} dependence on $\alpha_R$
is in the exponential factor, which we can determine. This allows us
to compare the dependence of the ratio of (mean-field) transition
temperatures on $\alpha_R$. For $\ell=0$ the equation $(\ref{eq:Vren
ell neq0 RG})$ holds as well, provided that we modify the initial
condition to
${g^{r}}^{(\ell=0)}_{\mu\lambda}|_{\Omega=A}=\frac{u}{4}\mu\lambda\sqrt{N_{\mu}N_{\lambda}}+\frac{1}{2^5}u^2m \sqrt{N_{\mu}N_{\lambda}}V^{(\ell=0)}_{\mu\lambda}$,
and use the eigenvalues of this matrix in the Eq.(\ref{eq:gren
flow}).

\begin{figure}[t]
\begin{center}
\includegraphics[width=0.475\textwidth]{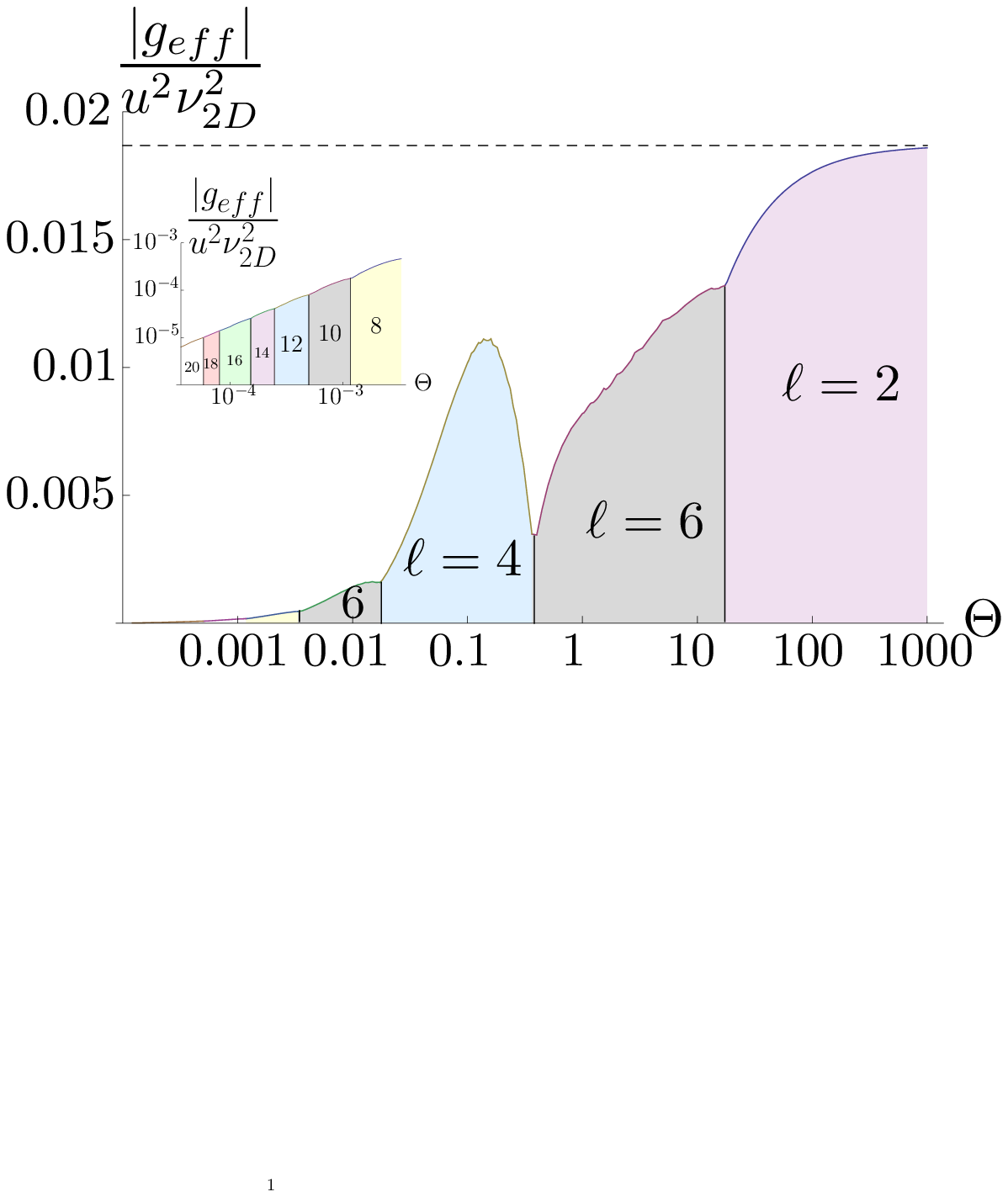}
\end{center}
  \caption{The effective coupling appearing in the expression for
  $T_c\approx Ae^{-1/|g_{eff}|}$ as a function of $\Theta=\frac{1}{2}m\alpha^2_R/E_F$. $\nu_{2D}=\frac{m}{2\pi}$. The dashed
  line at $0.0187$ is the $\Theta\rightarrow\infty$ asymptote.
  }\label{fig:phase diag}
\end{figure}

To within our numerical accuracy, we find that
$V^{(\ell=0)}_{--}=\frac{4}{\pi}$, while $V^{(\ell\neq0)}_{--}=0$,
for {\it any} $\Theta$. In addition, for $\Theta\gtrsim
\mathcal{O}(0.01)$ most dominant angle dependence is in $V_{++}$,
while there is only very weak angle dependence in $V_{+-}<0$. To
$\mathcal{O}(u)$, $g^{(\ell=0)}_+>0$, meaning no pairing
instability, and $g^{(\ell=0)}_-=0$. To $\mathcal{O}(u^2)$ we find that $g^{(\ell=0)}_->0$
for any $\Theta>0$, due to increase in {\it both}
$V^{(\ell=0)}_{++}$ and $V^{(\ell=0)}_{+-}$, latter of which becomes
less negative. This means that superconductivity resides
predominantly on the large Fermi surface and is
determined by some $V^{(\ell)}_{++}$ turning negative (meaning we
select $-$ in Eq.(\ref{eq:gell pm})). In Fig.\ref{fig:phase diag} we
show the $\Theta$ dependence of the couplings for the
$g_-^{(\ell)}$-channel which has the highest $T_c$. At small value
of $\Theta$, $\ell$ is very high (see inset of Fig.\ref{fig:phase
diag}). For the intermediate values of $\Theta$, starting with
$\sim0.005$, we find the sequence $\ell=6,4,6,2$, the last value of
which continues to $\Theta\rightarrow \infty$.

Finally, we need to determine which linear combination of the two
possible $\pm \ell$ states has the lowest (most negative)
condensation energy as we go below $T_c$. Adopting the arguments of
Anderson and Morel\cite{AndersonMorel1961}, we study this problem
below $T_c$ within mean-field. We replace the full angular
dependence of the pairing potential with just its projection on the
most dominant $\ell$ channel, an approximation which we expect to
hold away from the boundaries separating ground states with
different angular momentum. The self-consistent mean-field equations
are then solved near $T_c$ and at $T=0$. We find either a solution which
breaks time reversal symmetry and fully gaps the Fermi surface(s),
i.e. only one of the two $\pm\ell$ pairing components is finite, or
a solution with equal admixture of $\pm\ell$ and with gap nodes.
Comparing their condensation energies we find that the time reversal
breaking solution is lower by a factor of $1.5$ just below $T_c$ and by $e/2\approx 1.36$ at $T=0$. For
values of $\Theta \gtrsim 0.005$, the gap on the larger Fermi
surface is much larger than the gap on the smaller one due to the
smallness of ratio of $V_{+-}/V_{++}$. For smaller value of $\Theta$
the two gaps may be comparable.

In summary, we have studied the superconducting instability of a 2D
repulsive Fermi gas with Rashba spin-orbit coupling. We find that
due to the polarizable fermion background, the repulsion turns into
attraction on the large Fermi surface but not on the small one,
giving rise to pairing there. Additional Josephson tunneling,
$V^{(\ell)}_{+-}$, induces pairing on the small Fermi surface by
(weak) proximity effect. The resulting unconventional
superconducting states are found to break time reversal symmetry.
While the transition temperature is not strictly monotonic in the
dimensionless ratio $\Theta=\frac{1}{2}m\alpha_R^2/E_F$, the general
trend is that it grows with increasing $\Theta$. This experimentally
falsifiable feature, may provide means for enhancement of
superconductivity in a larger class of 2D electron systems.

Acknowledgements: We wish to thank Prof. L. P. Gor'kov for useful
discussions. This work is supported in part by NSF CAREER award
under Grant No. DMR-0955561.
\bibliography{SObib}
\end{document}